\documentclass[8pt]{article}
\usepackage{latexsym}
\usepackage{amssymb}
\usepackage{amsmath}
\usepackage{amscd}
\usepackage{amsthm}
\usepackage[left=2cm,top=2.5cm,right=2.5cm,bottom=1.5cm]{geometry}

\usepackage[dvips]{graphicx}
\usepackage{hyperref}

\begin{document}
\begin{center}
\Large{\bf{Some Bianchi Type-V Models of Accelerating Universe\\
with Dark Energy }}
\\
%\vspace{10mm} \normalsize{} \vspace{5mm}
\vspace{10mm} \normalsize{Suresh Kumar$^\dag$ and Anil Kumar Yadav$^\ddag$}\\ \vspace{4mm} \normalsize{
$^\dag$Department of Applied Mathematics, Delhi Technological University
(Formerly Delhi College of Engineering), Bawana Road, Delhi-110
042, India.}\\
\normalsize{E-mail: sukuyd@gmail.com}\\
\vspace{2mm}
\normalsize{$^\ddag$Department of Physics, Anand Engineering
College, Keetham, Agra-282 007, India} \\
\normalsize{E-mail: abanilyadav@yahoo.co.in, akyadav@imsc.res.in}\\
\end{center}
%%\date{}
%%\maketitle
\begin{abstract}
The paper deals  with a spatially homogeneous and anisotropic
universe filled with perfect fluid and dark energy components. The
two sources are assumed to interact minimally together with a
special law of variation for the average Hubble's parameter in order
to solve the Einstein's field equations. The law yields two explicit
forms of the scale factor governing the Bianchi-V space-time and
constant values of deceleration parameter. The role of dark energy
with variable equation of state parameter has been studied in detail
in the evolution of Bianchi-V universe. It has been found that dark
energy dominates the Universe at the present epoch, which is
consistent with the observations. The Universe achieves flatness
after the dominance of dark energy. The physical behavior of the
Universe has been discussed in detail.
\end{abstract}

Keywords: Bianchi-V space-time, Hubble's parameter, Deceleration
parameter, Dark energy.\\

PACS number: 98.80.Cq, 04.20.-q, 04.20.Jb
\smallskip
\vspace{4mm}
%%%%%%%%%%%%%%%%%%%%%%%%%%%%%%%%%%%%%%%%%%%%%%%%%%%%%%%%%%%%%%%%%%%%%%%%%%%%%%%%%%%%%%%%%%%%%%%%%%%%%
%%%%%%%%%%%%%%%%%%%%%%%%%%%%%%%   SECTION 1  %%%%%%%%%%%%%%%%%%%%%%%%%%%%%%%%%%%%%%%%%%%%%%%%%%%%%%%%%%
\section{Introduction}
Recent observations like type Ia supernova (SN Ia) \cite{ref1}$-$\cite{ref5}, CMB anisotropy \cite{ref6,ref7},
and large scale structure \cite{ref8} strongly indicate that our Universe is
spatially flat and there exists an exotic cosmic fluid called dark
energy (DE) with negative pressure, which constitutes about 70
percent of the total energy of Universe. It is an irony of the
nature and is a puzzling phenomenon that most abundant form of
matter-energy in the Universe is most mysterious. Many cosmologists
believe that the simplest candidate for the DE is the cosmological
constant ($\Lambda$) or vacuum energy since it fits the
observational data well. During the cosmological evolution, the
$\Lambda$-term has the constant energy density and pressure
$p^{(de)}=-\rho^{(de)}$. However, one has the reason to dislike the
cosmological constant since it always suffers from the theoretical
problems such as the ``fine-tuning" and ``cosmic coincidence"
puzzles \cite{ref9}. That is why, the different forms of
dynamically changing DE with an effective equation of state (EoS),
$\omega^{(de)}=p^{(de)}/\rho^{(de)}<-1/3$, have been proposed in the
literature. Other possible forms of DE include quintessence
($\omega^{(de)}>-1$)\cite{ref10}, phantom ($\omega^{(de)}<-1$)
\cite{ref11} etc. While the possibility $\omega^{(de)}<<-1$ is ruled
out by current cosmological data from SNe Ia (Supernovae Legacy
Survey, Gold sample of Hubble Space Telescope) \cite{ref5,ref12},
CMBR (WMAP, BOOMERANG) \cite{ref13,ref14} and large scale structure
(Sloan Digital Sky Survey) \cite{ref15} data, the dynamically
evolving DE crossing the phantom divide line (PDL)
$(\omega^{(de)}=-1)$ is mildly favored. Some other limits obtained
from the observational results coming from SN Ia data (Knop et al
\cite{ref16}) collaborated with CMBR anisotropy and galaxy
clustering statistics (Tegmark et al \cite{ref17}) are
$-1.67<\omega^{(de)}<-0.62$ and $-1.33<\omega^{(de)}<-0.79$
respectively.

Following Berman \cite{ref18} and Kumar and Singh \cite{ref19},
recently Singh et al. \cite {ref20} proposed a special law of
variation for the average Hubble's parameter in Bianchi-V
space-time, which yields a constant value of deceleration parameter.
Such a law of variation for Hubble's parameter is not inconsistent
with the observations and is also approximately valid for slowly
time-varying DP models. The law provides explicit forms of scale
factors governing the  Bianchi-V Universe and facilitates to
describe accelerating as well as decelerating modes of evolution of
the Universe. Models with constant DP have been extensively studied
in the literature in different contexts (see, Kumar and Singh
\cite{ref19} and references therein). Most of the models with
constant DP have been studied by considering perfect fluid or
ordinary matter in the Universe. But the ordinary matter is not
enough to describe the dynamics of an accelerating Universe. This
motivates the researchers to consider the models of the Universe
filled with some exotic matter such as the DE along with the usual
perfect fluid. Akarsu and Kilinc \cite{ref21}$-$\cite{ref23} have
investigated Bianchi-I and Bianchi-III DE models with constant DP.
Yadav and Yadav \cite{ref24} have studied the role of DE with
variable EoS in Bianchi type-III Universe evolving with constant DP.
Kumar \cite{ref25,ref26} has studied some isotropic and anisotropic
models of accelerating Universe with DE and constant DP. Recently,
Yadav et al \cite{ref27} has presented LRS Bianchi-V Universe with
DE characterized by variable EoS assuming constant DP.

In this paper, we have considered minimally interacting perfect
fluid and DE energy components with constant DP within the framework
of a Bianchi-V space-time in general relativity. The paper is
organized as follows. In Section 2, the models and field equations
have been presented. The Section 3 deals with the exact solutions of
the field equations and physical behavior of the models. Finally,
the results are discussed in section 4.

\section{Model and field equations}

The spatially homogeneous and anisotropic Bianchi-V space-time is
described by the line element
\begin{equation}\label{eq1}
ds^{2} =-dt^{2} +A^{2}dx^{2} +e^{2\alpha x}(B^{2}dy^{2}
+C^{2}dz^{2}),
\end{equation}
where  $A$ ,  $B$ and  $C$ are the metric functions of cosmic time
$t$ and $\alpha$ is a constant.

We define $a=(ABC)^{\frac{1}{3} } $ as the average scale factor of
the space-time (\ref{eq1}) so that the average Hubble's parameter
reads as
\begin{equation}\label{eq2}
H=\frac{\dot{a}}{a} ,
\end{equation}
where an over dot denotes derivative with respect to the cosmic time
$t$.

The Einstein's field equations in case of a mixture of perfect fluid
and DE components, in the units $8\pi G=c=1$,  read as
\begin{equation}\label{eq3}
R_{ij}-\frac{1}{2} g_{ij}R =- T_{ij},
\end{equation}
where $T_{ij}=T^{(m)}_{ij}+T^{(de)}_{ij}$ is the overall energy
momentum tensor with $T^{(m)}_{ij}$ and $T^{(de)}_{ij}$ as the
energy momentum tensors of ordinary matter and DE, respectively.
These are given by
\begin{equation}\label{eq4}
T^{(m)~i}_{~j}=diag[-\rho^{(m)},p^{(m)}\;,p^{(m)},p^{(m)}]\\
                   =diag[-1,\omega^{(m)},\omega^{(m)},\omega^{(m)}]\rho^{(m)}
\end{equation}
and
\begin{equation}\label{eq5}
T^{(de)~i}_{~j}=diag[-\rho^{(de)},p^{(de)},p^{(de)},p^{(de)}]\\
                   =diag[-1,\omega^{(de)},\omega^{(de)},\omega^{(de)}]\rho^{(de)}
\end{equation}
where $\rho^{(m)}$ and $p^{(m)}$ are, respectively the energy
density and pressure of the perfect fluid component or ordinary
baryonic matter while $\omega^{(m)}=p^{(m)}/\rho^{(m)}$ is its EoS
parameter. Similarly,  $\rho^{(de)}$ and $p^{(de)}$ are,
respectively the energy density and pressure of the DE component
while $\omega^{(de)}=p^{(de)}/\rho^{(de)}$ is the corresponding EoS
parameter.

In a comoving coordinate system, the field equations (\ref{eq3}),
for the Bianchi-I space-time (\ref{eq1}), in case of (\ref{eq4}) and
(\ref{eq5}), read as
\begin{equation}\label{eq6}
\frac{\ddot{B}}{B} +\frac{\ddot{C}}{C}
+\frac{\dot{B}\dot{C}}{BC}-\frac{\alpha^{2}}{A^{2}}=-\omega^{(m)}\rho^{(m)}-\omega^{(de)}\rho^{(de)},
\end{equation}
\begin{equation}\label{eq7}
\frac{\ddot{C}}{C} +\frac{\ddot{A}}{A}
+\frac{\dot{C}\dot{A}}{CA}-\frac{\alpha^{2}}{A^{2}}
=-\omega^{(m)}\rho^{(m)}-\omega^{(de)}\rho^{(de)},
\end{equation}
\begin{equation}\label{eq8}
\frac{\ddot{A}}{A} +\frac{\ddot{B}}{B}
+\frac{\dot{A}\dot{B}}{AB}-\frac{\alpha^{2}}{A^{2}}
=-\omega^{(m)}\rho^{(m)}-\omega^{(de)}\rho^{(de)},
\end{equation}
\begin{equation}\label{eq9}
\frac{\dot{A}\dot{B}}{AB} +\frac{\dot{B}\dot{C}}{BC}
+\frac{\dot{C}\dot{A}}{CA}-\frac{3\alpha^{2}}{A^{2}}
 =\rho^{(m)}+\rho^{(de)}.
\end{equation}
\begin{equation}\label{eq10}
2\frac{\dot{A}}{A}-\frac{\dot{B}}{B}
-\frac{\dot{C}}{C}=0
\end{equation}
The energy conservation equation $T^{(de)~ ij}_{~ ;j} =0$ yields
\begin{equation}\label{eq11}
\dot\rho^{(m)}+3(1+\omega^{(m)})\rho^{(m)}H+\dot\rho^{(de)}+3(1+\omega^{(de)})\rho^{(de)}H=0.
\end{equation}

\section{Solution of Field Equations}
Integrating (\ref{eq10}) and absorbing the constant of integration in
$B$ or $C$, without loss of generality, we obtain
\begin{equation}\label{eq12}
A^{2} =BC.
\end{equation}

Subtracting (\ref{eq6}) from (\ref{eq7}), (\ref{eq6}) from
(\ref{eq8}), (\ref{eq7}) from (\ref{eq8}) and taking second integral
of each, we get the following three relations respectively:
\begin{equation}\label{eq13}
\frac{A}{B} =d_{1} \exp \left(x_{1} \int a^{-3} dt \right),
\end{equation}
\begin{equation}\label{eq14}
\frac{A}{C} =d_{2} \exp \left(x_{2} \int a^{-3} dt \right),
\end{equation}
\begin{equation}\label{eq15}
\frac{B}{C} =d_{3} \exp \left(x_{3} \int a^{-3} dt \right),
\end{equation}
where  $d_{1} $ , $x_{1} $ ,  $d_{2} $ ,  $x_{2} $ ,  $d_{3} $  and
$x_{3} $ are constants of integration.

From equations (\ref{eq13})-(\ref{eq15}) and (\ref{eq12}), the metric functions can
be explicitly written as
\begin{equation}\label{eq16}
A(t)= a ,
\end{equation}
\begin{equation}\label{eq17}
B(t)=m a \exp \left(l \int a^{-3} dt \right),
\end{equation}
\begin{equation}\label{eq18}
C(t)=m^{-1} a \exp \left(-l \int a^{-3} dt \right).
\end{equation}
where
\begin{equation}
m =\sqrt[{3}]{d_{2} d_{3} } ,~~  l =\frac{(x_{2} +x_{3} )}{3}
\nonumber
\end{equation}
with
\begin{equation}\label{eq19}
d_{2}=d_{1}^{-1} ,~~  x_{2} =-x_{1}.
\end{equation}

In order to solve the field equations completely, first we assume
that the perfect fluid and DE components interact minimally.
Therefore, the energy momentum tensors of the two sources may be
conserved separately.

The energy conservation equation $T^{(m)~ij}_{~;j} =0$, of the
perfect fluid leads to
\begin{equation}\label{eq20}
\dot\rho^{(m)}+3(1+\omega^{(m)})\rho^{(m)}H=0,
\end{equation}
whereas the energy conservation equation
$T^{(de)~ij}_{~;j} =0$, of the DE component
yields
\begin{equation}\label{eq21}
\dot\rho^{(de)}+3(1+\omega^{(de)})\rho^{(de)}H=0.
\end{equation}

Following Akarsu and Kilinc \cite{ref21}, we assume that the EoS
parameter of the perfect fluid to be a constant, that is,
\begin{equation}\label{eq22}
\omega^{(m)}=\frac{p^{(m)}}{\rho^{(m)}}=const.,
\end{equation}
while $\omega^{(de)}$ has been allowed to be a function of time
since the current cosmological data from SNIa, CMB and large scale
structures mildly favor dynamically evolving DE crossing the PDL as
discussed in Section 1.

Now integration of equation (\ref{eq20}) leads to
\begin{equation}\label{eq23}
\rho^{(m)}=c_{0}a^{-3(1+\omega^{(m)})},
\end{equation}
where $c_{0}$ is a positive constant of integration.

Finally, we constrain the system of equations with a law of
variation for the average Hubble parameter in Bianchi-V space-time
proposed by Singh et al. \cite{ref20}, which yields a constant value
of DP. The law reads as
\begin{equation}\label{eq24}
H=Da^{-n},
\end{equation}
where  $D>0$  and  $n\geq 0$ are constants.  In the following
subsections, we discuss the DE cosmology for $n\neq0$ and $n=0$ by
using the law (\ref{eq24}).

\subsection{DE Cosmology for $n\neq 0$}
In this case, integration of (\ref{eq24}) leads to
\begin{equation}\label{eq25}
a(t)=(nDt+c_{1})^{\frac{1}{n}},
\end{equation}
where $c_{1}$ is a constant of integration.

Inserting (\ref{eq25}) into (\ref{eq16})-(\ref{eq18}), we get
\begin{equation}\label{eq26}
A(t)=(nDt+c_{1} )^{\frac{1}{n} } ,
\end{equation}
\begin{equation}\label{eq27}
B(t)=m (nDt+c_{1} )^{\frac{1}{n} } \exp \left[\frac{l }{D(n-3)}
(nDt+c_{1} )^{\frac{n-3}{n} } \right] ,
\end{equation}
\begin{equation}\label{eq28}
C(t)=m^{-1} (nDt+c_{1} )^{\frac{1}{n} } \exp \left[\frac{-l
}{D(n-3)} (nDt+c_{1} )^{\frac{n-3}{n} } \right] ,
\end{equation}

provided $n\neq3$.

Therefore, the model (\ref{eq1}) becomes
\begin{equation}\label{eq29}
ds^{2} =-dt^{2} +T^{\frac{2}{n}}\left(dx^{2} +m^{2}e^{2\alpha
x+kT^{\frac{n-3}{n}}}dy^{2} +m^{-2}e^{2\alpha
x-kT^{\frac{n-3}{n}}}dz^{2}\right),
\end{equation}
where $T=nDt+C_{1}$ and $k=\frac{2l}{D(n-3)}$.

The average Hubble's parameter ($H$), energy density ($\rho^{(m)}$)
of perfect fluid, DE density ($\rho^{(de)}$) and EoS parameter
($\omega^{(de)}$) of DE, for the model (\ref{eq29}) are found to be
\begin{equation}\label{eq30}
H=D(nDt+c_{1})^{-1},
\end{equation}
\begin{equation}\label{eq31}
\rho^{(m)}=c_{0}(nDt+c_{1})^{-\frac{3(1+\omega^{(m)})}{n}},
\end{equation}
\begin{equation}\label{eq32}
\rho^{(de)}=3D^{2}(nDt+c_{1})^{-2}-l^{2}(nDt+c_{1})^{-\frac{6}{n}}
-3\alpha^{2}(nDt+c_{1})^{-\frac{2}{n}}-c_{0}(nDt+c_{1})^{-\frac{3(1+\omega^{(m)})}{n}},
\end{equation}
\begin{equation}\label{eq33}
\omega^{(de)}=\frac{(2n-3)D^{2}(nDt+c_{1})^{-2}-l^{2}(nDt+c_{1})^{-\frac{6}{n}}
+\alpha^{2}(nDt+c_{1})^{-\frac{2}{n}}-c_{0}\omega^{(m)}(nDt+c_{1})^{-\frac{3(1+\omega^{(m)})}{n}}}{3D^{2}(nDt+c_{1})^{-2}-l^{2}(nDt+c_{1})^{-\frac{6}{n}}
-3\alpha^{2}(nDt+c_{1})^{-\frac{2}{n}}-c_{0}(nDt+c_{1})^{-\frac{3(1+\omega^{(m)})}{n}}}.
\end{equation}

The above solutions satisfy the equation (\ref{eq29}) identically,
as expected.

The spatial volume ($V$) and expansion scalar $(\theta)$ of the
model read as
\begin{equation}\label{eq34}
V=a^{3}=(nDt+c_{1})^{\frac{3}{n}},
\end{equation}
\begin{equation}\label{eq35}
\theta=3H=3D(nDt+c_{1})^{-1}.
\end{equation}

The anisotropy parameter ($\bar{A}$) and shear scalar $(\sigma)$ of
the model are given by
\begin{equation}\label{eq36}
\bar{A}=\frac{1}{9H^{2}}\left[\left(\frac{\dot{A}}{A}-\frac{\dot{B}}{B}\right)^{2}
+\left(\frac{\dot{B}}{B}-\frac{\dot{C}}{C}\right)^{2}
+\left(\frac{\dot{C}}{C}-\frac{\dot{A}}{A}\right)^{2}\right]
=\frac{2l^{2}}{3D^{2}}(nDt+c_{1})^{-\frac{2(3-n)}{n}},
\end{equation}
\begin{equation}\label{eq37}
\sigma^{2}=\frac{3}{2}\bar{A}H^{2}=l^{2}(nDt+c_{1})^{\frac{-6}{n}}.
\end{equation}
The value of DP ($q$) is found to be
\begin{equation}\label{eq38}
q=-\frac{a\ddot{a}}{{\dot{a}}^{2}} = n-1,
\end{equation}
which is a constant. The sign of $q$ indicates whether the model
inflates or not. A positive sign of $q$, i.e., $n>1$ corresponds to
the standard decelerating model whereas the negative sign of $q$,
i.e., $0< n<1$ indicates inflation. The expansion of the Universe at
a constant rate corresponds to $n=1$, i.e., $q=0$. Also, recent
observations of SN Ia \cite{ref1}$-$\cite{ref5} reveal that the
present Universe is accelerating and value of DP lies somewhere in
the range $-1<q< 0.$ It follows that in the derived model, one can
choose the values of DP consistent with the observations.

\begin{figure}
\begin{center}
\includegraphics [height=6 cm]{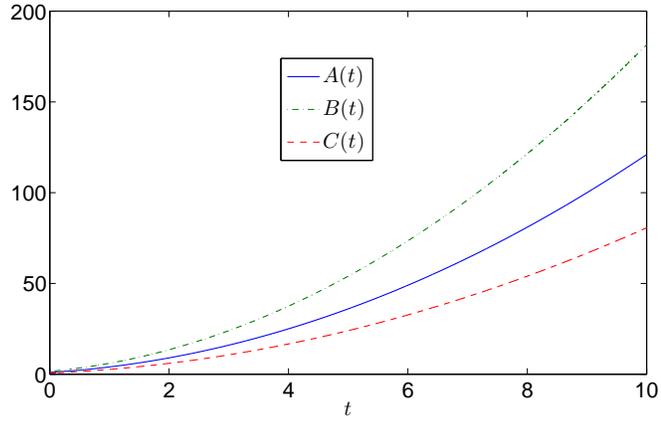}
\caption{Scale factors vs. time with $m=1.5$, $l=0.3$, $D=2$,
$c_{1}=1$, $n=0.5$.} \label{fg:skakF5.eps}
\end{center}
\end{figure}

\begin{figure}
\begin{center}
\includegraphics [height=6 cm]{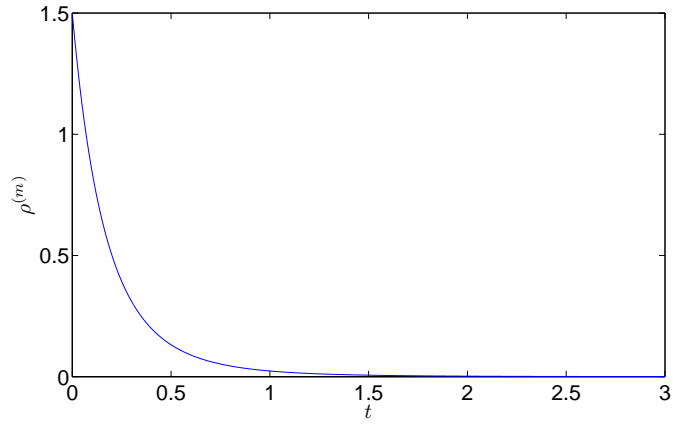}
\caption{Matter energy density vs. time with $c_{0}=1.5$, $c_{1}=1$,
$D=2$, $n=0.5$, $\omega^{(m)}=0$.} \label{fg:skakF1.eps}
\end{center}
\end{figure}

\begin{figure}
\begin{center}
\includegraphics[height=6 cm]{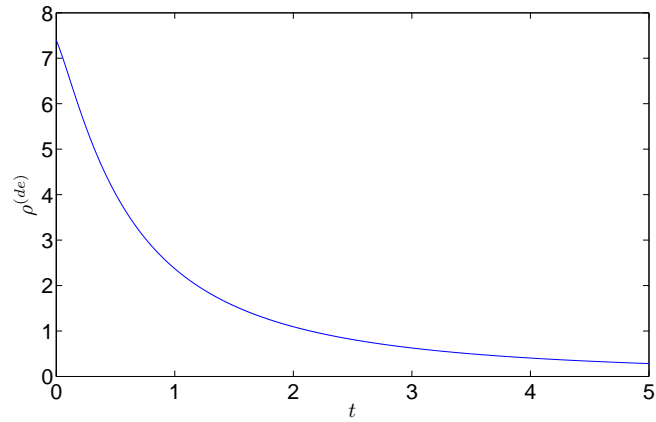}
\caption{DE density vs. time with $c_{0}=1.5$, $c_{1}=1$, $D=2$,
$n=0.5$, $\omega^{(m)}=0$, $l=1.2$, $\alpha=0.7$.}
\label{fg:skakF2.eps}
\end{center}
\end{figure}
We observe that at $t=-c_{1}/nD$, the spatial volume vanishes while
all other parameters diverge. Therefore, the model has a big bang
singularity at $t=-c_{1}/nD$, which can be shifted to $t=0$ by
choosing $c_{1}=0$. The singularity is point type as the directional
scale factors $A(t)$, $B(t)$ and $C(t)$ vanish at the initial
moment.

\textbf{Fig.1} shows that the directional scale factors
monotonically increase with time. Thus, expansion of the Universe
takes place in all the three spatial directions.

From \textbf{Fig. 2} and \textbf{Fig. 3}, we observe that
$\rho^{(m)}$ as well as $\rho^{(de)}$ remain positive during the
cosmic evolution. Therefore, the weak energy condition (WEC) as well
as null energy condition (NEC) are obeyed in the derived model.
Further, $\rho^{(m)}$ and $\rho^{(de)}$ decrease with time, and
approach to small positive values at the present epoch.

The parameters $H$, $\theta$ and $\sigma^{2}$ start off with
extremely large values, and continue to decrease with the expansion
of the Universe. The anisotropy parameter $\bar{A}$ also decreases
with the cosmic evolution provided $n<3$. This shows that anisotropy
of the model goes off during the cosmic evolution. The spatial
volume grows with the cosmic time.

\begin{figure}
\begin{center}
\includegraphics[height=6 cm]{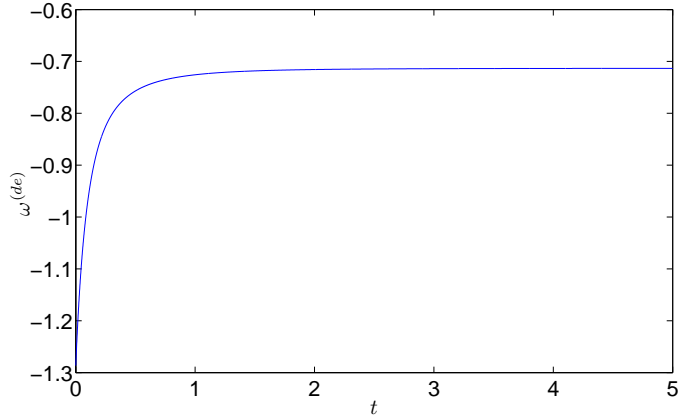}
\caption{EoS parameter of DE vs. time with $c_{0}=1$, $c_{1}=0.05$,
$D=2$, $n=0.5$, $\omega^{(m)}=0$, $l=1.2$, $\alpha=0.7$.}
\label{fg:skakF3.eps}
\end{center}
\end{figure}

\begin{figure}
\begin{center}
\includegraphics[height=6 cm]{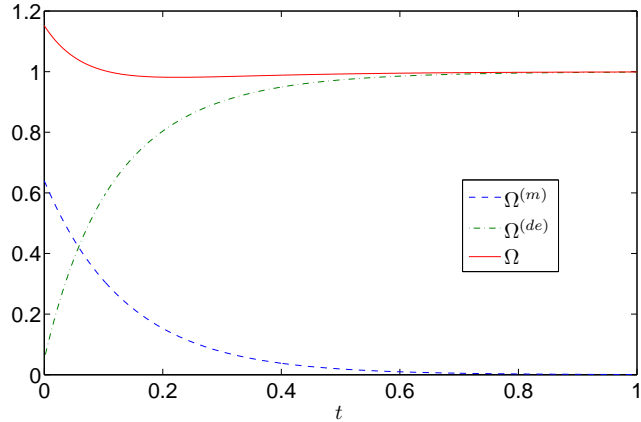}
\caption{Density parameters vs. time with $D=2.5$, $c_{0}=12$,
$c_{1}=1$,
 $\omega^{(m)}=0$, $n=0.05$, $l=1.2$, $\alpha=0.7$.}
\label{fg:skakF4.eps}
\end{center}
\end{figure}

\textbf{Fig. 4} depicts the variation of EoS parameter
$(\omega^{(de)})$ versus cosmic time for accelerating phase of
Universe ($q=-0.5$), as a representative case with appropriate
choice of constants of integration and other physical parameters .
The SN Ia data suggests that $-1.67<\omega^{(de)}<-0.62$
\cite{ref16} while the limit imposed on $\omega^{(de)}$ by a
combination of SN Ia data (with CMB anisotropy) and galaxy
clustering statistics is $-1.33<\omega^{(de)}<-0.79$ \cite{ref17}.
\textbf{Fig. 4}, clearly shows that $\omega^{(de)}$ evolves within a
range, which is in nice agreement with SN Ia and CMB observations.

The perfect fluid density parameter $(\Omega^{(m)})$ and DE density
parameter $(\Omega^{(de)})$ are given by
\begin{equation}
\label{eq39}
\Omega^{(m)}=\frac{c_{0}}{3D^2}(nDt+c_{1})^{\frac{2n-(1+\omega^{(m)})}{n}},
\end{equation}
\begin{equation}
\label{eq40}
\Omega^{(de)}=1-\frac{l^2}{3D^2}(nDt+c_{1})^{\frac{2n-6}{n}}-\frac{\alpha^2}{D^2}(nDt+c_{1})^{\frac{2n-2}{n}}
-\frac{c_{0}}{3D^2}(nDt+c_{1})^{\frac{2n-(1+\omega^{(m)})}{n}}.
\end{equation}

The overall density parameter $(\Omega)$ is given by
\begin{equation}
\label{eq41} \Omega=\Omega^{(m)} + \Omega^{(de)} =
1-\frac{l^2}{3D^2}(nDt+c_{1})^{\frac{2n-6}{n}}-
\frac{\alpha^2}{D^2}(nDt+c_{1})^{\frac{2n-2}{n}}.
\end{equation}

From equation $(\ref{eq41})$, it is observed that for $0 < n < 1$ ,
the overall density parameter $(\Omega)$ approaches to 1 at late
times. Thus, the derived model predicts a flat Universe at the
present epoch. {\bf Fig. 5}, depicts the variation of density
parameters versus cosmic time during the evolution of Universe. It
is observed that initially Universe was matter dominated and later
on DE dominates the evolution of Universe which is probably
responsible for the accelerated expansion of present-day Universe.
\subsection{DE Cosmology for $n=0$}
In this case, integration of (\ref{eq24}) yields
\begin{equation}\label{eq42}
a(t)=c_{2}e^{Dt},
\end{equation}
where $c_{2}$ is a positive constant of integration.

The metric functions, therefore, read as
\begin{equation}\label{eq43}
A(t)=c_{2}e^{Dt},
\end{equation}
\begin{equation}\label{eq44}
B(t)=m c_{2} \exp \left(Dt-\frac{l }{3Dc_{2}^{3} } e^{-3Dt} \right),
\end{equation}
\begin{equation}\label{eq45}
C(t)=m^{-1} c_{2} \exp \left(Dt+\frac{l }{3Dc_{2}^{3} } e^{-3Dt}
\right).
\end{equation}
Therefore, the model (\ref{eq1}) becomes
\begin{equation}\label{eq46}
ds^{2}=-dt^{2}+c_{2}^{2}e^{2Dt}\left(dx^{2} +m^{2}e^{2\alpha
x+hle^{-3Dt}}dy^{2} +m^{-2}e^{2\alpha x-hle^{-3Dt}}dz^{2}\right).
\end{equation}
where $h=\frac{-2 }{3Dc_{2}^{3} }$.

The average Hubble's parameter, energy density of perfect fluid, DE
density and EoS parameter of DE, for the model (\ref{eq46}) are
obtained as
\begin{equation}\label{eq47}
H=D,
\end{equation}
\begin{equation}\label{eq48}
\rho^{(m)}=c_{0}c_{2}^{-3(1+\omega^{(m)})}e^{-3D(1+\omega^{(m)})t},
\end{equation}
\begin{equation}\label{eq49}
\rho^{(de)}=3D^{2}-l^{2}c_{2}^{-6}e^{-6Dt}-3\alpha^{2}c_{2}^{-2}e^{-2Dt}
-c_{0}c_{2}^{-3(1+\omega^{(m)})}e^{-3D(1+\omega^{(m)})t},
\end{equation}
\begin{equation}\label{eq50}
\omega^{(de)}=\frac{-3D^{2}-l^{2}c_{2}^{-6}e^{-6Dt}+\alpha^{2}c_{2}^{-2}e^{-2Dt}}{3D^{2}-l^{2}c_{2}^{-6}e^{-6Dt}-3\alpha^{2}c_{2}^{-2}e^{-2Dt}
-c_{0}c_{2}^{-3(1+\omega^{(m)})}e^{-3D(1+\omega^{(m)})t}}.
\end{equation}

The above solutions satisfy the equation (\ref{eq46}) identically,
as expected.

The spatial volume and expansion scalar of the model read as
\begin{equation}\label{eq51}
V=c_{2}^{3}e^{3Dt},
\end{equation}
\begin{equation}\label{eq52}
\theta=3D.
\end{equation}

The DP is given by
\begin{equation}\label{eq53}
q=-1.
\end{equation}

The perfect fluid density parameter and DE density parameter are
given by
\begin{equation}
\label{eq54}
\Omega^{(m)}=\frac{c_{0}c_{2}^{-3(1+\omega^{(m)})}}{3D^2}e^{-3D(1+\omega^{(m)})t},
\end{equation}
\begin{equation}
\label{eq55}
\Omega^{(de)}=1-\frac{l^{2}c_{2}^{-6}}{3D^{2}}e^{-6Dt}-\frac{\alpha^{2}c_{2}^{-2}}{D^{2}}e^{-2Dt}
-\frac{c_{0}c_{2}^{-3(1+\omega^{(m)})}}{3D^{2}}e^{-3D(1+\omega^{(m)})t}.
\end{equation}

The overall density parameter reads as
\begin{equation}
\label{eq56} \Omega =
1-\frac{l^{2}c_{2}^{-6}}{3D^{2}}e^{-6Dt}-\frac{\alpha^{2}c_{2}^{-2}}{D^{2}}e^{-2Dt}.
\end{equation}

The anisotropic parameter and shear scalar of model (\ref{eq46}) are
given by
\begin{equation}
\label{eq58}
 \bar{A}=\frac{2l^{2}c_{2}^{-6}}{3D^2}e^{-6Dt},
\end{equation}
\begin{equation}
\label{eq59} \sigma^2=l^{2}c_{2}^{-6}e^{-6Dt}.
\end{equation}

Recent observations of SN Ia \cite{ref1}$-$\cite{ref5}, suggest that
the Universe is accelerating in its present state of evolution. It
is believed that the way Universe is accelerating presently; it will
expand at the fastest possible rate in future and forever. For
$n=0$, we get \textbf{ $q=-1$ }; incidentally this value of DP leads
to $dH/dt=0$, which implies the greatest value of Hubble's parameter
and the fastest rate of expansion of the Universe. Therefore, the
derived model can be utilized to describe the dynamics of the late
time evolution of the actual Universe. So, in what follows, we
emphasize upon the late time behavior of the derived model.

From Equation (\ref{eq50}), we observe that $\omega^{(de)}\approx
-1$ for sufficiently large time $t$. Therefore, the late time
dynamics of EoS parameter $\omega^{(de)}$ represents the vacuum
fluid dominated Universe, which is mathematically equivalent to
cosmological constant. Further, at late times, we have\\

\indent $\rho^{(m)}\approx0$,\\

$\rho^{(de)}\approx3D^2$,\\

$\bar{A}\approx0$.\\

Thus, the ordinary matter density becomes negligible whereas the
accelerated expansion of Universe continues with non-zero and
constant DE density, which is consistent with recent observations.
Also the late time dynamics of the derived model shows that the
anisotropy of the Universe damps out and Universe becomes isotropic.
From equation (\ref{eq56}), it is observed that for sufficiently
large time, the overall density parameter approaches to 1, i.e.,
$\Omega\approx 1$. Thus, the model predicts a flat Universe.

%%%%%%%%%%%%%%%%%%%%%%%%%%%%%%%%%%%%%%%%%%%%%%%%%%%%%%%%%%%%%%%%%%%%%%%
%%%%%%%%%%%%%%%%%%%%%%%%%%  SECTION 4  %%%%%%%%%%%%%%%%%%%%%%%%%%%%%%%%%%
\section{Concluding Remarks}
In this paper, we have investigated the role of DE with variable EoS
parameter in the evolution of Universe within the framework of a
spatially homogeneous and anisotropic Bianchi-V space-time by taking
into account the special law of variation of average Hubble
parameter, which yields a constant value of DP. The main features of
the work are as follows:

\begin{itemize}
    \item The models are based on exact solutions of the field
    equations.
    \item In the present models, the
matter energy density and DE density remain positive. Therefore, the
weak and null energy conditions are satisfied, which in turn imply
that the derived models are physically realistic.

    \item The singular model ($n\neq 0$) seems to describe the dynamics of
    Universe from big bang to the present epoch while the
    non-singular model ($n=0$) seems reasonable to project dynamics of
    future Universe.

    \item The age of Universe, in the singular model, is given by
$$T_{0} = \left(\frac{1}{D}\right)H_{0}^{-1}$$
which differs from the present estimate, i.e., $T_{0} = H_{0}^{-1}
\approx 14Gyr$. But if we take $D = 1$, the model $(n\neq 0)$ is in
good agreement with the present age of Universe.

 \item The EoS parameter of DE evolves within the range predicted by the observations.

    \item The DE dominates the Universe at the present epoch, which
    may be attributed to the current accelerated expansion of the
    Universe.

    \item The Universe acquires flatness with the dominance of DE (see,
    \textbf{Fig.5}).
    Thus, flatness of the Universe seems a natural consequence
    of DE.

\end{itemize}

Hypothetical DE is the most popular way of explaining why the
Universe is expanding at an ever increasing rate. DE plays a massive
part in shaping our reality however nobody seems certain of what the
dang stuff actually is. Future space missions hope to solve this
mystery and shake up our current understanding of the Universe.
%\newline
%\nonumsection{References}

\end{document}